\documentclass[11pt]{article}
\usepackage{epsfig}
\usepackage{amsfonts}
\usepackage{amsmath}
\usepackage{bbm}
\usepackage{bm}
\usepackage{color}
\usepackage{slashed}
\usepackage{cite}

\usepackage[colorlinks=false,citecolor=cyan,urlcolor=blue,bookmarks=true,bookmarks=true,bookmarksopen=true,bookmarksnumbered=true,bookmarksopenlevel=3]{hyperref}

\definecolor{airforceblue}{rgb}{0.36, 0.54, 0.66}
\definecolor{steelblue}{rgb}{0.27, 0.51, 0.71}
\definecolor{amber}{rgb}{1.0, 0.49, 0.0}

\allowdisplaybreaks[1]
\def\simg{{\ \lower-1.2pt\vbox{\hbox{\rlap{$>$}\lower6pt\vbox{\hbox{$\sim$}}}}\ }}
\def\siml{{\ \lower-1.2pt\vbox{\hbox{\rlap{$<$}\lower6pt\vbox{\hbox{$\sim$}}}}\ }}

\makeatletter \@addtoreset{equation}{section} \makeatother

\begin{document}

\flushbottom

\begin{titlepage}

\begin{centering}

\vfill

{\Large{\bf
  Leptogenesis and composite heavy neutrinos with gauge mediated interactions 
}} 

\vspace{0.8cm}

S.~Biondini$^{\rm a}$ and 
O.~Panella$^{\rm b}$ 

\vspace{0.8cm}

{\em
$^{\rm a}$AEC, Institute for Theoretical Physics, 
University of Bern, \\ 
Sidlerstrasse 5, CH-3012 Bern, Switzerland\\
$^{\rm b}$Istituto Nazionale di Fisica Nucleare, Sezione di Perugia
\\
Via  A.  Pascoli,  I-06123  Perugia,  Italy} 

\vspace*{0.8cm}

\end{centering}

\vspace*{0.3cm}
 
\noindent
\textbf{Abstract}: Leptogenesis is an appealing framework to account for the baryon asymmetry in the universe. To this end  physics beyond the Standard Model is demanded. In this paper we investigate the possibility to attain successful leptogenesis with composite Majorana neutrinos. We work in the framework of effective gauge mediated and contact interactions without any reference to an underlying  compositeness theory. This approach is the one adopted in all current experimental searches for composite fermions at colliders.  In the case of gauge mediated interactions, we calculate the CP asymmetry in heavy composite neutrino decays. Both the direct and indirect CP asymmetry are derived and resonant leptogenesis is also discussed. We find that the Sakharov conditions can be met and, for some choice of the parameters, the correct order of magnitude of the baryon asymmetry is reproduced.

\vfill

\vfill
\newpage
\tableofcontents
\end{titlepage}

\section{Introduction}
\setcounter{page}{3}
There are strong evidences that the standard model (SM) alone cannot provide any exhaustive solution to the dark matter problem and the baryon asymmetry in the universe. On the dark matter side the only candidates that can be provided within the SM are neutrinos. Due to their tiny masses, those particles can possibly act as hot dark matter and hence be responsible at most for a very small fraction of the required dark matter energy density. As far as the generation of a baryon asymmetry (or \textit{baryogenesis}) is concerned, the Sakharov conditions \cite{Sakharov:1967dj} are fulfilled by the SM. However the CP violation is too small and the departure from thermal equilibrium is not strong enough so that physics beyond the SM appears necessary \cite{Gavela:1994ds,Jarlskog:1985ht,Gavela:1994dt,Kajantie:1995kf}. 

In this respect it is important to scrutinize the available models, that extend the SM particle spectrum to address fundamental questions in particle physics, in view of their possible connection with the striking observations from astrophysics and cosmology. It happens quite often that one can borrow new fields and interactions to have a suitable dark matter candidate or a successful mechanism for a matter-antimatter asymmetry generation. On the other hand the parameter space of the model at hand can be constrained in order to satisfy the accurate measurements of dark-matter and baryon-asymmetry abundances in the universe. Indeed it sounds a win-win situation. For example, a very well-motivated, constrained and predictive model is the neutrino minimal SM ($\nu$MSM) \cite{Asaka:2005pn}. Here three right-handed sterile neutrinos with masses below the electroweak scale are responsible for the active neutrino masses via the seesaw mechanism and, in some regions of the parameter space, the model can account for both dark matter and the the observed baryon asymmetry via leptogenesis. We remind to extensive reviews for more details \cite{Drewes:2013gca, Adhikari:2016bei}. Of course supersymmetry  provides suitable dark matter candidates depending on the choice of the lightest supersymmetric particle \cite{Ellis:2010kf}, together with heavy Majorana fermions that make baryogenesis working thorough leptogenesis \cite{Plumacher:1997ru, Giudice:2003jh}. Axions may work as dark matter particles as well and they originally arise to solve the strong CP problem \cite{Duffy:2009ig, Kim:1986ax}.

In this paper we focus on the connection between models that contain heavy composite fermions, especially Majorana neutrinos, and the possibility to realize a successful baryogenesis via leptogenesis\cite{Fukugita:1986hr}. Compositeness is one of the many  scenarios of physics beyond the SM on the market \cite{Dimopoulos:1980hn,Fritzsch:1981zh,Pati:1974yy,Terazawa:1976xx,Neeman:1979wp,Shupe:1979fv,Harari:1980ez, Barbieri:1981cy}. Composite models for quark and leptons are usually advocated to address the proliferation of elementary fermions of the three generations, their mass and mixing patterns and their similar behaviour under the electroweak interactions. In this approach ordinary fermions are regarded as composite objects of unknown more fundamental constituents (sometimes called \textit{preons}). Despite it is quite hard to build a quantum field theory of fundamental interactions of such sub-constituents, many features of these models can be addressed phenomenologically and are indeed subject to experimental searches. 

The common ground of composite models is to assume a high energy scale, usually denoted with $\Lambda$, below which composite fermions interact effectively among themselves via contact interactions~\cite{Eichten:1983hw,Baur:1989kv}. The more fundamental degrees of freedom are not resolved and the interactions are accounted for dimension-six operators suppressed by the high energy scale $\Lambda$. Moreover the possibility to have excited quarks and leptons is in order. The mass scale of these heavier resonances is often denoted with $M^*$ and they can interact among themselves and with SM fermions via contact and gauge interactions. We call them heavy composite states throughout the paper. Gauge mediated interactions are an alternative, and complementary, way to implement models for composite fermions~\cite{Cabibbo:1983bk, Pancheri:1984sm}. In this case, the quantum numbers of the composite fermions can be fixed by the weak isospin invariance, similarly to when  strong isospin invariance was used to fix the characteristics of many (unknown) hadronic states later experimentally observed. Among the excited leptons, electromagnetically neutral states are found which can be accommodated in a rather simple way to be Majorana fermions.

Direct searches of composite fermions have been performed at colliders for quite some time. 
To date the stronger bounds are provided by the LHC experiments. Updated bounds
on charged excited lepton masses  
have been provided by the LHC Run I analyses, where ATLAS and CMS collaborations give respectively a lower bound $M^* >2.2$ TeV \cite{Aad:2013jja} and $M^*>2.45$ TeV ($M^*>2.48$ TeV) \cite{Khachatryan:2015scf} for heavy composite electrons (muons). Those bounds are extracted imposing $\Lambda=M^*$, moreover contact interactions are used for the production of the heavy composite leptons and gauge interactions to account for their decays. A phenomenological driven study has recently investigated the accessible parameter space for heavy composite neutrinos at the LHC, where both gauge and contact interactions have been included in the production cross sections and decays leading to a like-sign dilepton plus di-jet final state signature \cite{Leonardi:2015qna}. A dedicated experimental analysis by the CMS collaboration excludes heavy-composite neutrinos with masses $M^*<4.35 (4.70)$ TeV~\cite{CMSexp,Sirunyan:2017xnz} for a di-jet$\,ee$ (di-jet$\,\mu \mu$) final state, when $M^*=\Lambda$.  

Our aim is to inspect possible connections between composite heavy neutrinos and leptogenesis, however, we want to keep the discussion as much as possible model independent. Therefore we take the effective Lagrangians for both contact and gauge mediated interactions (see eqs.~(\ref{lag_mirr})-(\ref{cont_2}) below), without any reference to an underlying theory. In doing so, we allow for more direct comparison with the experimental constraints on the model parameters driven by the very same Lagrangians. 

In order to assess the leptogenesis mechanism within a given model, one needs to check at least the three Sakharov conditions: lepton number violation (LNV), C and CP violation, out-of-equilibrium dynamics. In particular we consider composite models that accommodate Majorana neutrinos and we calculate the corresponding composite-neutrino decay widths into SM leptons and antileptons. Those are the key ingredients for the generation of a lepton asymmetry if one assumes complex couplings that add new sources of CP violation.  
Leptogenesis with composite neutrinos has been investigated in \cite{Grossman:2008xb}, where the authors consider an underlying SU(6)$_c$ preon dynamics as a particular case of confining gauge theories presented in \cite{Dimopoulos:1980hn}. Here the composite heavy neutrinos are responsible for the smallness of the SM neutrino masses via the seesaw mechanisms, and for leptogenesis at the same time.  Bounds on the heavy composite neutrinos is of the order of $10^{10}$ GeV, therefore out of range with respect to collider searches. Recently another study has been carried out in \cite{Zhuridov:2016xls} where composite heavy neutrinos, labelled as leptomesons and arising from the UV completion proposed in \cite{Fritzsch:1981zh}, are responsible for the generation of the matter-antimatter asymmetry. In this case four-fermion contact interactions are considered and both leptogenesis from heavy-particle decays and oscillations is addressed. The CP asymmetry coming from the interference between the tree-level and the one-loop wave-function diagram has been considered (often referred to as indirect CP asymmetry), mainly in the case of a resonant enhancement. Moreover, giving up the simultaneous explanation of SM neutrino masses and the baryon asymmetry, composite states with mass scale of the TeV scale can be pursued. 

The structure of the paper is as follows: in section~\ref{sec_model} we introduce the model of composite fermions that comprises Majorana neutrinos and the corresponding effective Lagrangians, together with 
the discussion of the necessary conditions for leptogenesis. 
In section~\ref{sec_width} we calculate explicitly the widths al leading order induced by the gauge-mediated interactions and show their complementarity in the parameter space $(\Lambda, M^*)$ with contact-interaction induced widths. In the case of gauge interactions, we provide  the expressions for the CP violating parameters that require the evaluation of two-loop cut diagrams. We consider and calculate both the indirect and direct CP asymmetry, and discuss two limits for the composite neutrino mass spectrum, namely a strongly hierarchical and nearly degenerate spectrum. All these results are discussed in section~\ref{sec_CP}. The out-of-equilibrium dynamics is addressed in section~\ref{sec_out} and conclusions are found in section~\ref{sec_conc}.  

\section{Composite-neutrino models and leptogenesis}
\label{sec_model}
The Majorana or Dirac nature of neutrinos is a longstanding puzzle in contemporary physics. The implication of a Majorana mass term for neutrinos may have a big impact on the processes occurring in the early universe. In particular Majorana fermions are a key ingredient for baryogenesis via leptogenesis. A Majorana  mass term automatically leads to lepton-number-violating (LNV) processes because one cannot assign a definite lepton charge to a Majorana fermion. The violation of lepton number is the first condition to be fulfilled for a successful leptogenesis, together with CP violation and the out-of-equilibrium dynamics. These three requirements are the so-called Sakharov conditions \cite{Sakharov:1967dj} and we shall discuss them in the context of composite neutrino models in this section.

In this paper we consider standard thermal leptogenesis induced by heavy particle decays. We do not deal with leptogenesis via neutrino oscillations~\cite{Akhmedov:1998qx}. In the former scenario, heavy Majorana fermions are at the origin of a lepton-asymmetry induced by their lepton number and CP violating decays in SM leptons/antileptons. We briefly recall the simplest leptogenesis mechanism with heavy Majorana neutrinos \cite{Fukugita:1986hr,Buchmuller:2004nz}. The heavy states are kept in equilibrium with the plasma at sufficiently high temperatures by decay, inverse decay and scattering processes. However, the expansion and then cooling of the universe makes the temperature dropping below the heavy neutrino mass. Then the heavy states effectively decay into SM particles and produce a different amount of leptons and antileptons due to the CP violating phases that distinguish matter from antimatter. Such net imbalance is not washed out by the inverse decays that are Boltzmann suppressed (and scatterings inefficient as well). 
It is important to remark that thermal leptogenesis has to occur at temperatures above the electroweak phase transition. This makes sure that any lepton asymmetry is partially reprocessed into a baryon asymmetry through the sphaleron transitions in the SM~\cite{Kuzmin:1985mm}.\footnote{Actually the temperature at which the sphaleron transitions switch off is the relevant one. This is not exactly the same of the electroweak crossover temperature, respectively they are $T_{sph} \approx 130$ GeV and $T_{c} \approx 160$ GeV, see~\cite{Laine:2015kra,DOnofrio:2014rug} for recent precision studies.} We stick to this standard scenario and hence we work in an unbroken phase of the SM  gauge group, SU(2)$_L \times$U(1)$_Y$, in the following. 

We propose that heavy composite neutrinos interacting with gauge and contact interactions  may lead to a successful leptogenesis. A fundamental assumption is in order. The typical temperatures during the onset of leptogenesis and the effective composite-neutrino decays have to be smaller than the compositeness scale, $\Lambda$. On the contrary the composite states would dissolve in the sub-constituents and the description in terms of composite Majorana fermions would not be viable. The mass scale of heavy-composite neutrinos is not subjected to the constrained imposed by the seesaw mechanism in this model, at least in its realization discussed in the present paper, so we assume it to be of order of the TeV scale.  
\subsection{Gauge and contact interaction Lagrangians}
There are at least two possibilities to accommodate Majorana composite neutrinos when considering effective gauge-mediated interactions: the \textit{sequential-type} and the \textit{mirror-type } model~\cite{Hagiwara:1985wt,Takasugi:1995bb,Olivepdg}. The first option comprises excited states whose left-handed  components  are  accommodated in a SU(2) doublet whereas the right-handed components are SU(2) singlets. Following the discussion in \cite{Takasugi:1995bb,Olivepdg}, if the right-handed excited neutrino is not considered, only a Majorana mass term can be obtained and this determines the Majorana nature of the fermion. 
The second option, the mirror-type model,  contains an excited right-handed doublet and left-handed singlets. Again we can assume that the left-handed excited neutrino is absent. Therefore we can associate a Majorana mass term to the composite neutrino. In this paper we consider only the mirror-type model for the gauge interactions.\footnote{At colliders the cross sections and widths induced by the sequential model are expected to be suppressed by a factor $v/\Lambda$, where $v$ is the vacuum expectation value of the Higgs field, with respect to the mirror model.} The corresponding gauge mediated Lagrangian reads
\begin{equation}
\mathcal{L}_{{\hbox{\tiny mir}}}= \frac{1}{2\Lambda} \bar{L}_{L,\alpha}  \sigma^{\mu \nu}  \left( f \, g  \tau^a W^{a}_{\mu \nu} + f' \, g' \frac{Y}{2} B_{\mu \nu} \right)  L_{R,I}^* + h. c. \, ,
\label{lag_mirr}
\end{equation} 
where $L^T_{L ,\alpha}=(\nu_{L,\alpha}, e_{L,\alpha})$ is a SM SU(2)$_L$ doublet with flavour $\alpha=e,\mu,\tau$, then $\sigma^{\mu \nu} = i [\gamma^\mu , \gamma^\nu]/2$, $g$ and $g'$ are the SU(2)$_L$ and U(1)$_Y$ gauge couplings respectively, $f$ and $f'$ the effective couplings of the model, $W_{\mu \nu}$ and $B_{\mu \nu}$ are the field strength for the SU(2)$_L$ and U(1)$_Y$ gauge fields, $\tau^a=\sigma^a/2$ where $\sigma^a$ are the  Pauli matrices, $Y$ is the hypercharge of the doublets, and $L^{*T}_{R,I}=(\nu^*_{R,I}, e^*_{R,I})$ stands for the composite lepton doublet ($I$ is understood as mass eigenstate index). Such Lagrangian describes the interaction between a left-handed SM doublet and a right-handed composite doublet
mediated by the SM gauge fields. 
A comment is on order: in principle one can consider the excited left(right)-handed neutrino in the mirror(sequential) model. In this case one may write a Dirac mass term in addition to the Majorana mass term and implement a seesaw mechanism in the heavy neutrino sector. However we stick to the choice as introduced in \cite{Takasugi:1995bb,Olivepdg} in order  to keep a simpler realization of the model for the following discussion.  

Another way to implement effective interactions between excited and SM fermions is with four-particle contact interactions. They can be understood as arising from constituent exchanges that are not resolved at energies smaller than $\Lambda$. The corresponding Lagrangian comprises two fermion currents and hence two inverse powers of the high energy scale $\Lambda$ appear. It reads 
\begin{equation}
\mathcal{L}_{{\hbox{\tiny cont}}}=\frac{g_*^2}{2\Lambda^2} j^{\mu}j_{\mu} \, ,
\label{cont_1}
\end{equation}   
where the vector current is
\begin{eqnarray}
j^{\mu}=\eta_L \bar{\psi}_L \gamma^{\mu} \psi_L +\eta'_L \bar{\psi}^*_L \gamma^{\mu} \psi^*_L  +\eta''_L \bar{\psi}^*_L \gamma^{\mu} \psi_L + h.c. + (L \to R) \, ,
\label{cont_2}
\end{eqnarray}
where $g^2_*=4 \pi$ and the $\eta$'s are constants of order one. They are put equal to one in the literature when doing phenomenological and experimental studies, as well as the couplings $f$ and $f'$. Moreover only chirality in the current (\ref{cont_2}) is considered in most of the phenomenological and experimental studies and we adopt the same choice in the following. We retain the right-handed chirality in (\ref{cont_2}) to build the Majorana mass term.
\subsection{Composite models and Sakharov conditions}
We now come to discuss in detail the Sakharov conditions. First, by extending the SM sector with either the gauge mediated or contact interactions and Majorana composite neutrinos, lepton number violation is introduced. Second, we replace the $\eta$'s, $f=\pm f'$~\cite{Olivepdg} with  complex couplings, $\tilde{\eta}_{\alpha I}$ and $\tilde{f}_{\alpha I}$ respectively, where $\alpha$ stands for the flavour index of the SM lepton and $I$ for the mass index of the composite neutrino (in the literature these couplings are usually taken as real).  Whereas for the model in (\ref{lag_mirr})  such a complexification is straightforward, this is not the case for the model that comprises contact interactions in (\ref{cont_1}). Indeed we can introduce complex couplings only for the operators in (\ref{cont_2}) which are
not self-adjoint, namely the operators multiplied by $\eta''_L$ ($\eta''_R$), within
the context of a CPT-invariant theory. The complex couplings are responsible for additional CP violating phases with respect to the SM ones. The corresponding Lagrangian reads, in basis where the composite neutrino mass matrix is diagonal and expressing them with Majorana fields $N^*_I=\nu^*_{I,R} + (\nu^*_{I,R})^c$, as follows
\begin{eqnarray}
\mathcal{L}_{\hbox{\tiny gauge}}&=&\mathcal{L}_{\hbox{\tiny SM}}+\frac{1}{2} \bar{N}^*_I i \slashed{\partial} N^*_I - \frac{M_I}{2} \bar{N}^*_I N_I^* + \frac{g}{\sqrt{2} \Lambda} \left[ \tilde{f}_{\alpha I} \, \bar{e}_\alpha \sigma^{\mu \nu} \partial_\mu W^-_\nu P_R N^{*}_I + h.c. \right]\nonumber
\\
&+& \frac{\tilde{g}}{2 \Lambda} \left[ \tilde{f}_{\alpha I} \, \bar{\nu}_\alpha \sigma^{\mu \nu} \partial_\mu Z_\nu P_R N^{*}_I + h.c. \right]  + \dots \, ,
\label{eff_lag_gau}
\end{eqnarray} 
where dots stand for terms which are not relevant in the following (comprising the charged-composite lepton), then $\tilde{g}=\sqrt{g^2+g'^2}$ and $\mathcal{L}_{\hbox{\tiny SM}}$ is the SM Lagrangian with an unbroken gauge group (massless fermions and gauge bosons). The notation for the SM leptons is as follows: $\nu_\alpha$ for a neutral lepton and $e_{\alpha}$ for a charged lepton, with $\alpha=e,\mu,\tau$. The appearance of the physical $Z$ boson is due to an assumption we made, namely $f=f'$. According to this choice one obtains the field combination $\tilde{g} Z_\mu \equiv g W^3_\mu -g'B_\mu$. In this case there is no interaction between composite neutrinos and photons. On the other hand one may take $f=-f'$ and then the field combination $\tilde{g} \bar{Z}_\mu \equiv g W^3_\mu +g'B_\mu$ would enter in (\ref{eff_lag_gau}), which can be seen as a linear superposition of the physical eigenstates, the $Z$ boson and the photon. Such  assumption on $f$ and $f'$ may be dropped and the Lagrangian would comprise more involved interactions with the fields $B_{\mu}$ and $W_{\mu}^3$. However for illustration, and in order not to introduce many different complex couplings, we assume $f= f' \to \tilde{f}_{\alpha I} $. 

For contact interactions the Lagrangian is
\begin{equation}
\mathcal{L}_{\hbox{\tiny contact}}=\mathcal{L}_{\hbox{\tiny SM}}+\frac{1}{2} \bar{N}^*_I i \slashed{\partial} N^*_I - \frac{M_I}{2} \bar{N}^*_I N_I^* +\frac{g_*^2}{2\Lambda^2} \left[ \tilde{\eta}_{\alpha I} \bar{\psi} \gamma_{\mu} P_R \psi'  \, \bar{\ell}_\alpha \gamma^{\mu} P_R N^*_I  + h.c. \right] + \dots \, ,
\label{eff_lag_cont}
\end{equation}
where the SM lepton $\ell_{\alpha}$ is either neutral, $\nu_{\alpha}$, or charged, $e_{\alpha}$, and $\bar{\psi} \gamma_{\mu} \psi'$ stands for either a SM lepton or quark current. A remark here is in order: depending on $\ell_{\alpha}$ being a neutral or charged SM lepton, the accompanying SM  current is constrained to be a neutral or charged current accordingly to preserve electric charge. As far as the complexifiaction of the couplings is concerned, the assignment is $\tilde{\eta} \equiv \eta^{\phantom{*}}_R \eta''^{*}_R$, where the complex nature of $\tilde{\eta}$ is only induced by $\eta''_R$.

\section{Composite neutrino widths}
\label{sec_width}
In this section we derive the expressions for the composite-neutrino widths at order $\tilde{f}^2$ and $\tilde{\eta}^2$ in the Yukawa couplings. Our fundamental object is the two-point function of the composite heavy-neutrino field, that reads
\begin{equation} 
-i \left. \int d^{4}x \, e^{ip\cdot x} \, \langle
\Omega | T \left( N_{I}^{*\mu}(x) \bar{N}_{I}^{*\nu}(0) \right) | \Omega \rangle \right|_{p^\alpha =(M_I + i\epsilon,\bm{0}\,)} \, ,
\label{matrixFund}
\end{equation} 
where $| \Omega \rangle$ stands for the ground state of the fundamental theory. 
We compute corrections to the composite neutrino propagator. We are interested in the imaginary part of the corresponding loop diagrams that are related to a width according to the optical theorem. This may appear too technical for a leading order decay width, however it helps in setting the formalism for the computation of the CP asymmetry in the next section. Since we work in the unbroken electroweak phase all the SM particles are massless, on the contrary heavy composite neutrinos are massive due to an unknown underlying dynamics. 
\begin{figure}[t!]
\centering
\includegraphics[scale=0.5]{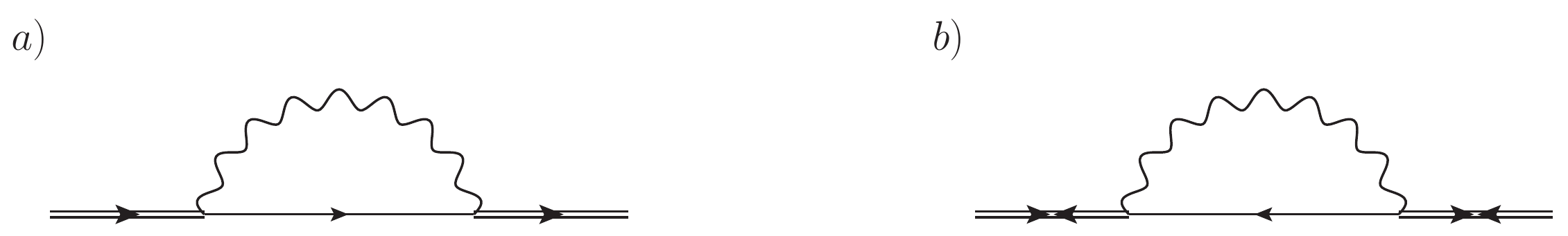}
\caption{\label{fig:fig1} One-loop self-energy diagrams for the composite neutrino. Solid lines stand for a SM lepton, either charged or neutral, and wiggled lines for the SM gauge bosons ($Z$ and $W^\pm$). Solid double lines stand for a composite neutrino: forward arrows correspond to the $\langle N^* \bar{N^*} \rangle$ contraction, whereas forward-backward arrows to $\langle N^* N^* \rangle$ and $\langle \bar{N^*} \bar{N^*} \rangle$, the latter being typical of a Majorana fermion.}
\end{figure}
We calculate the composite neutrino width in its rest frame and at zero temperature. Its momentum can be taken as $p^\mu=M v^{\mu}$ with $v^{\mu}=(0, \bm{1})$. In doing so the non-trivial Dirac algebra due to the magnetic coupling can be rather simplified due to non-relativistic projectors arising from heavy-neutrino external legs (see appendix in \cite{Biondini:2013xua}). Moreover the Majorana nature of the composite neutrino has to be accounted for by the different contractions of the Majorana spinors. This is shown in figure~\ref{fig:fig1} with a solid forward arrow and forward-backward arrow respectively, where the wiggly line stands for either $Z$ or $W^{\pm}$ gauge bosons. The leading order result reads
\begin{equation}
\Gamma_{I,\alpha}^{\hbox{\tiny gauge}}=\frac{(g'^2+3g^2)}{32 \pi } \left( \frac{M^*_I }{\Lambda} \right)^2  |\tilde{f}_{I\alpha}|^2 M^*_I\, ,
\label{gaugewidth_1}
\end{equation} 
where the superscript stands for the gauge-induced decay width and the subscripts for the lepton flavour in the final state and the composite-neutrino generation. Gauge invariance has been explicitly checked. The imaginary part of the loop amplitude is needed to obtain the width, or alternatively, one can use cutting rules \cite{Cutkosky:1960sp,Remiddi:1981hn,Denner:2014zga} to reduce it to a tree level computation with two particles in the final state.  

\begin{figure}[t!]
\centering
\includegraphics[scale=0.825]{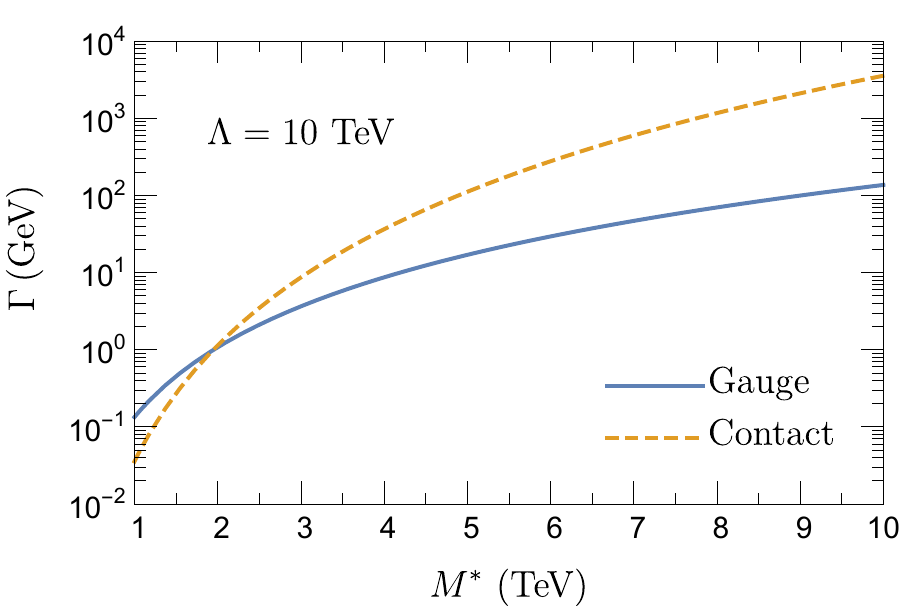}
\hspace{0.01 cm}
\includegraphics[scale=0.825]{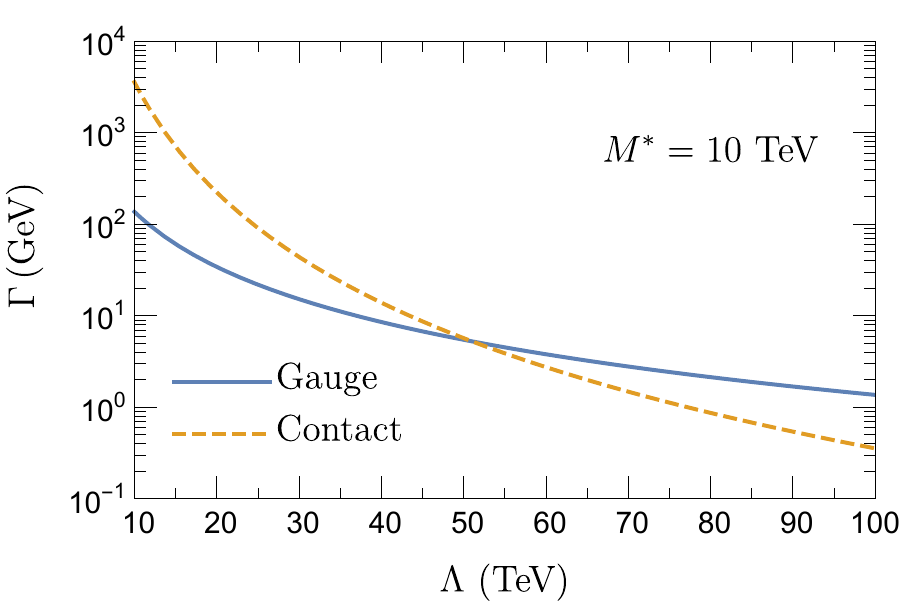}
\caption{\label{fig:fig2} The plot show gauge and contact-induced widths, in solid blue and dashed orange respectively. In the left panel the gauge and contact-induced widths are shown for masses spanning from 1 to 10 TeV, for a fixed value of $\Lambda=10$ TeV. In the right panel the same widths are shown for different values of $\Lambda$, assuming a composite neutrino mass $M^*=10$ TeV. The gauge couplings are evaluated at the scale $M^*$, whereas $|\tilde{f}_{I\alpha}|^2=|\tilde{\eta}_{I\alpha}|^2\equiv 1$.}
\end{figure}

Composite neutrino decays into a lepton/antilepton are also induced from the contact interaction Lagrangian (\ref{eff_lag_cont}), that comprises dimension-six operators.  At leading order we find
\begin{equation}
\Gamma_{I,\alpha}^{\hbox{\tiny contact}}= \frac{107}{1536}\frac{g_*^4}{ \pi^3 } \left( \frac{M^*_I }{\Lambda} \right)^4  |\tilde{\eta}_{I\alpha}|^2 M^*_I \,  \, .
\label{gaugecont_1}
\end{equation} 
In the case of contact interactions, the decay process is highly inclusive and many different combinations are comprised in the final state.\footnote{For $\nu_\alpha$ being a final state lepton in the composite-neutrino decay process, we have to consider neutral SM currents. On the other hand when a charged lepton $e_\alpha$ appears as a decay product, charged SM currents enter.} Our result in (\ref{gaugecont_1}) differs from that given in~\cite{Zhuridov:2016xls}, due to different final states allowed in the two distinct models. We find that the widths induced by the gauge and contact interactions are complementary in the parameter space.  Results are shown in figure~\ref{fig:fig2}. In the left panel we show the gauge- and contact-induced widths for masses spanning from 1 to 10 TeV, for a fixed value of $\Lambda=10$ TeV. Gauge-induced widths (solid blue line) dominate for $M^*  \siml 2$ TeV over the contact widths (dashed orange line). In the right panel the gauge and contact-induced widths for different values of $\Lambda$ are shown, assuming a composite neutrino mass $M^*=10$ TeV. The different power suppression $(M^*/ \Lambda)$ and couplings appearing in eqs.~(\ref{gaugewidth_1}) and (\ref{gaugecont_1}) are responsible for the relative importance of the two different widths. We remark that gauge (contact) interactions induce a two-body (three-body) decay of the composite neutrino.
The complementarity of gauge and contact induced widths in the model parameter space $(M^*,\Lambda)$ has been already noticed in studies related to the phenomenology at colliders in a broken SU(2)$_L \times$U(1)$_Y$ phase~\cite{Leonardi:2015qna}. However, there are some differences with the situation considered here. For gauge interactions, we do not need to consider further decays of the gauge bosons in the decay $N_{I} \to \ell_{\alpha} + \text{gauge boson}$. Such particle content is all what is needed to single out a LNV process and induce leptogenesis. For the contact interactions, we have to sum over many processes to consider the inclusive process $N^*_I \to \ell_{\alpha} + X$, where $X$ stands for any fermion-antifermion pair that does not carry a net lepton number.  In the present work we focus on composite neutrinos in the framework of gauge mediated interactions and inspect the corresponding CP asymmetries. This is the subject of the next section. We will treat in detail the CP asymmetries from contact interactions elsewhere~\cite{BP2}.
\begin{figure}[t!]
\centering
\includegraphics[scale=0.48]{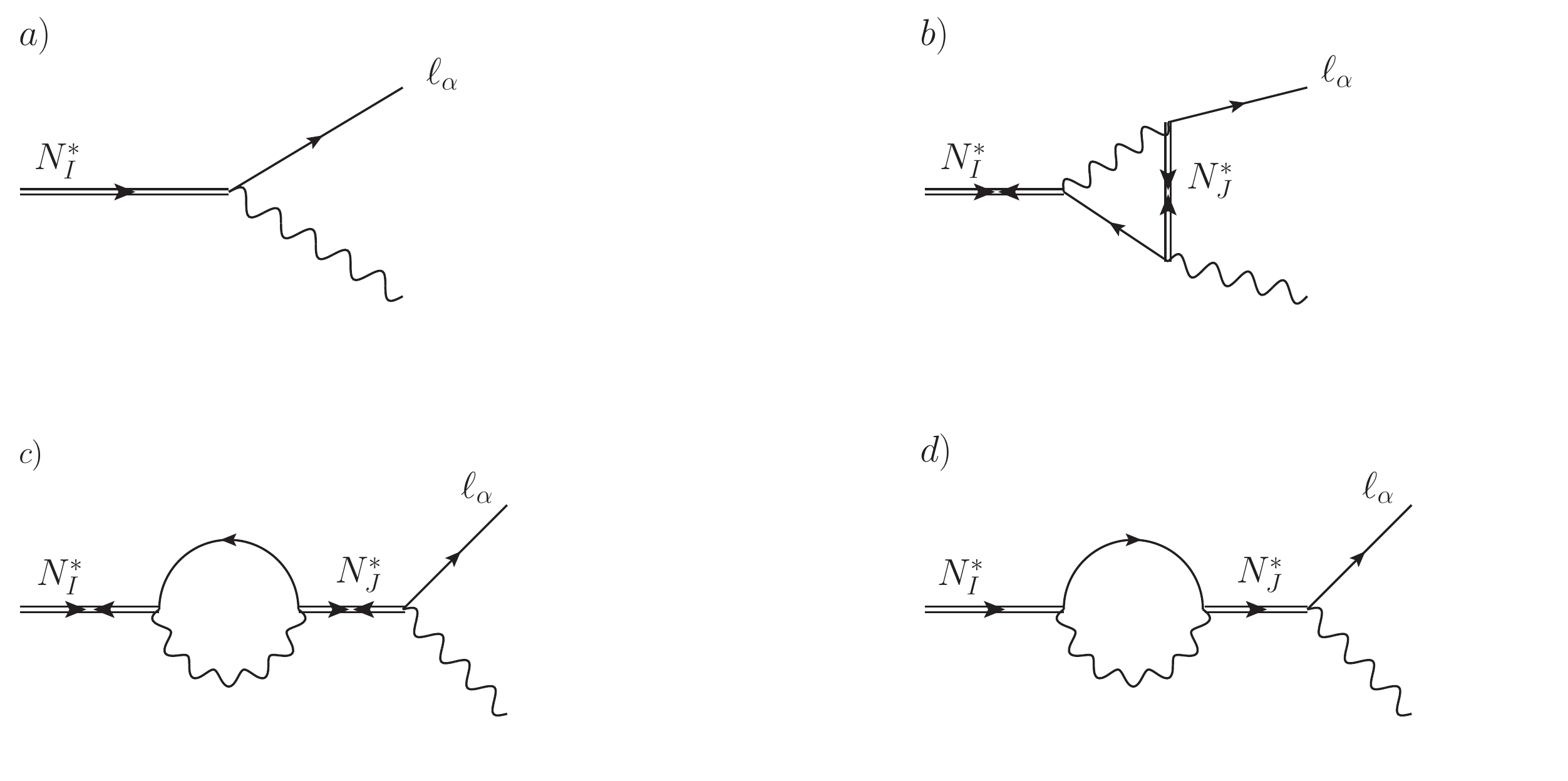}
\caption{\label{fig:fig3} CP asymmetries are originated from the interference between tree-level and one-loop vertex and self-energy (or wave-function) diagrams. 
Solid double lines stand for heavy composite neutrinos, solid lines for SM leptons and wiggled lines for gauge bosons. The neutrino propagator with forward and forward-backward arrows have the same meaning as in figure~\ref{fig:fig2}.}
\end{figure}   
\section{CP asymmetry}
\label{sec_CP}
According to standard thermal leptogenesis, heavy Majorana neutrinos populate the early universe with an equilibrium abundance at temperatures larger than their mass scale and then start to decay out-of-equilibrium when the temperature of the plasma drops below the heavy-particle mass. Indeed the back reaction, mainly inverse decays, are Boltzmann suppressed in such regime. Here we consider an analogous situation, so that heavy composite neutrinos can decay into SM leptons and antileptons in different amounts, due to the CP-violating phases in $\tilde{f}_{I \alpha}$. The CP parameter is defined as usual
\begin{equation}
\epsilon_{I,\alpha}=
 \frac{\Gamma(N^*_I \to \ell_{\alpha} + \text{gauge boson})-\Gamma(N^*_I \to \bar{\ell}_{\alpha}+ \text{gauge boson} )  }
{\sum_{\alpha} \Gamma(N^*_I  \to \ell_{\alpha} + \text{gauge boson} ) + \Gamma(N^*_I  \to \bar{\ell}_{\alpha}+ \text{gauge boson})} \, ,
\label{eq:adef}
\end{equation}
where the sum runs over the SM lepton flavours, $N^*_I$  stands for the
$I$-th heavy composite neutrino species, $\ell_{\alpha}$ is a SM lepton with flavour $\alpha$. We shall provide the flavoured CP asymmetries in the following. Indeed the unflavoured
regime is an appropriate choice at very high temperatures, namely $T \simg 10^{12}$ GeV~\cite{Nardi:2005hs,Nardi:2006fx}, on the contrary the three lepton flavours are resolved by the thermal bath and the CP asymmetries are stored in each flavour. 

\begin{figure}[t!]
\centering
\includegraphics[scale=0.56]{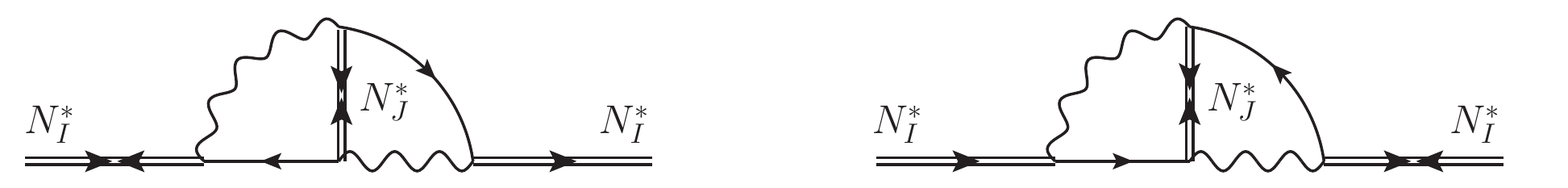}
\caption{\label{fig:fig4}Self-energy diagrams for the $I$-th composite Majorana neutrino. The imaginary parts, namely the cutting through lepton and gauge boson lines, corresponds to the interference of tree level and one-loop vertex diagram in figure~\ref{fig:fig3}. We show the two possible diagrams due to Majorana field contractions.}
\end{figure}

\begin{figure}[t!]
\centering
\includegraphics[scale=0.5]{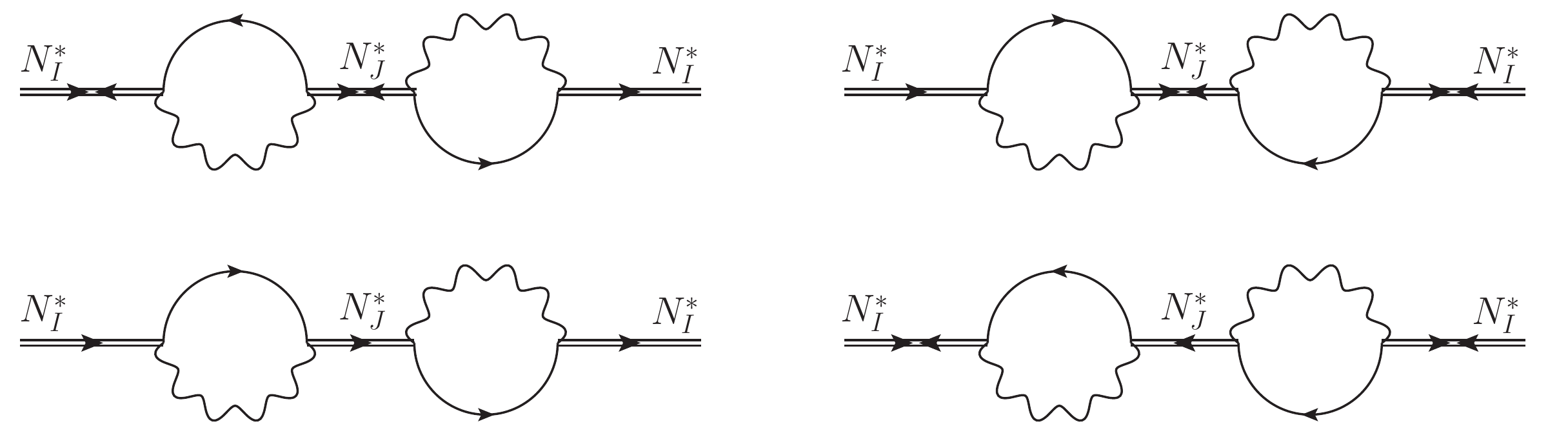}
\caption{\label{fig:fig5}Self-energy diagrams for the $I$-th composite Majorana neutrino. The imaginary parts, namely the cutting through lepton and gauge boson lines, corresponds to the interference of tree level and one-loop self-energy diagrams in figure~\ref{fig:fig3}. We show the two possible diagrams due to Majorana field contractions in each raw.}
\end{figure}

The CP asymmetry can be calculated  from the interference between the tree-level and one-loop diagrams \cite{Covi:1996wh,Davidson:2008bu}, and we show them in figure~\ref{fig:fig3} for the model under study. 
Diagram $b$ is referred to as the vertex diagram, whereas diagrams $c$ and $d$ are often called wave-function (or self-energy) diagrams. 
Moreover diagram $d$ is relevant only for the flavoured CP asymmetry because its contribution vanishes when summing over the lepton flavour in the final state.
In analogy with the standard leptogenesis case, we find that the contribution to the CP asymmetry depends on the composite-neutrino mass spectrum. In the literature two limits are often considered, the hierarchical and the nearly degenerate ones. First we compute the CP asymmetry for a mass spectrum $M^*_1 < M^*_2 < M^*_3$ and then we consider those limits of the mass pattern.
The interference between the tree-level and one-loop diagrams in figure~\ref{fig:fig3} may be computed 
from the imaginary part of the heavy-neutrino self-energy at order $\tilde{f}^4$. As mentioned in section~\ref{sec_width}, our fundamental object is the composite-neutrino self-energy and now we study the corresponding corrections at order $\tilde{f}^4$ instead of $\tilde{f}^2$. Therefore two-loop self-energy diagrams are considered and we show them in figure~\ref{fig:fig4} and~\ref{fig:fig5}. Moreover we are interested in singling out the contribution to the leptonic and antileptonic heavy-neutrino decays. This is necessary to keep track of the different decay rates into matter or antimatter according to (\ref{eq:adef}). 
Cutting rules are exploited to select a lepton or an antilepton as final state particle in the two-loop diagrams~\cite{Cutkosky:1960sp,Remiddi:1981hn,Denner:2014zga}. A detailed example of their implementation for diagrams with the same topology can be found in \cite{Biondini:2015gyw,Biondini:2016arl}, where standard leptogenesis with heavy Majorana neutrinos and seesaw type I is considered. 
Following the same notation, we may write the flavoured CP asymmetry \eqref{eq:adef} 
due to the vertex diagram, $\epsilon^{\hbox{\tiny direct}} _{I,\alpha} $,
and to the wave-function diagram, $\epsilon^{\hbox{\tiny indirect}} _{I,\alpha}$, in the general form 
\begin{eqnarray}
\epsilon_{I,\alpha}= \epsilon^{\hbox{\tiny direct}} _{I,\alpha} + \epsilon^{\hbox{\tiny indirect}} _{I,\alpha} = 
&&- 2 \sum_{J} {\rm{Im}}(B_{\hbox{\tiny direct}}+ B_{\hbox{\tiny indirect}}^{\hbox{\tiny LNV}})  \frac{ {\rm{Im}} \left[ (\tilde{f}_{I}^{*}\tilde{f}^{\phantom{*}}_{J})(\tilde{f}^*_{\alpha I}\tilde{f}^{\phantom{*}}_{\alpha J}) \right] }{|\tilde{f}_{I}|^2}
\nonumber
\\
&&- 2 \sum_{J} {\rm{Im}}(B^{\hbox{\tiny LNC}}_{\hbox{\tiny indirect}})  \frac{{\rm{Im}} \left[ (\tilde{f}^{\phantom{*}}_{I}\tilde{f}_{J}^{*})(\tilde{f}^*_{\alpha I}\tilde{f}^{\phantom{*}}_{\alpha J}) \right] }{|\tilde{f}_{I}|^2} ,
\label{CPdef2}
\end{eqnarray}
where  $(\tilde{f}_{I}^{*}\tilde{f}^{\phantom{*}}_{J}) \equiv \sum_\alpha \tilde{f}^*_{\alpha I}\tilde{f}^{\phantom{*}}_{\alpha J}$. For the indirect contribution, we separate the LNV and lepton number conserving (LNC) contributions explicitly in complete analogy with the case of standard leptogenesis~\cite{Covi:1996wh}.  
The functions $B_{\hbox{\tiny direct}}$, $B_{\hbox{\tiny indirect}}^{\hbox{\tiny LNV}}$ and $B_{\hbox{\tiny indirect}}^{\hbox{\tiny LNC}}$ can be calculated
by cutting the two-loop diagrams in figure~\ref{fig:fig3} and evaluating the remaining one-loop diagram. Then only the imaginary part of the one-loop vertex and wave-function diagrams is relevant and we find it to be finite.  We refer the reader to appendix~\ref{app1} for some details on the loop integrals.  
The result for the CP asymmetry from the vertex topology reads
\begin{eqnarray}
\epsilon_{I,\alpha}^{\hbox{\tiny direct}} &=& \frac{5g^4+2g^2g'^2+g'^4}{3g^2+g'^2}\sum_{J\neq I}  \left( \frac{M^*_I}{\Lambda}\right)^2  \frac{ {\rm{Im}}\left[ (\tilde{f}_{I}^{*}\tilde{f}^{\phantom{*}}_{J})(\tilde{f}^*_{\alpha I}\tilde{f}^{\phantom{*}}_{\alpha J})  \right] }{8\pi |\tilde{f}_I|^2} \mathcal{F} \left[ \left( \frac{M^*_J}{M^*_I}\right)^2 \right] \, , \nonumber \\
\label{cp_direct}
\end{eqnarray}
where  $\mathcal{F}(x)$ is defined, in a similar fashion as in \cite{Covi:1996wh}, as follows
\begin{eqnarray}
\mathcal{F}(x) = \frac{\sqrt{x}}{8} \left[ 1+2x-2x(1+x)  \ln\left(1+\frac{1}{x}\right)   \right] \, .
\label{direct_loop}
\end{eqnarray}
The indirect CP asymmetry comprises two different contributions and the result is
\begin{eqnarray}
\epsilon^{\hbox{\tiny indirect}}_{I,\alpha} &=& \frac{5g^4+2g^2g'^2+g'^4}{3g^2+g'^2} \sum_{J\neq I}  \left( \frac{M^*_I}{\Lambda}\right)^2 \left\lbrace  \frac{ {\rm{Im}}\left[ (\tilde{f}_{I}^{*}\tilde{f}^{\phantom{*}}_{J})(\tilde{f}^*_{\alpha I}\tilde{f}^{\phantom{*}}_{\alpha J})  \right] }{8 \pi |\tilde{f}_I|^2} \mathcal{G}^{\hbox{\tiny LNV}} \left[ \left( \frac{M^*_J}{M^*_I}\right)^2 \right]   \right.  \nonumber \\
  &+&  \left.  \frac{ {\rm{Im}}\left[ (\tilde{f}_{I}^{*}\tilde{f}^{\phantom{*}}_{J})(\tilde{f}^{\phantom{*}}_{\alpha I}\tilde{f}^*_{\alpha J})  \right] }{8 \pi |\tilde{f}_I|^2} \mathcal{G}^{\hbox{\tiny LNC}} \left[ \left( \frac{M^*_J}{M^*_I}\right)^2\right]   \right\rbrace \, ,
\label{cp_indirect}
\end{eqnarray}
where the loop-functions are respectively 
\begin{eqnarray}
&&\mathcal{G}^{\hbox{\tiny LNV}}(x)=\frac{\sqrt{x}}{8}\frac{1}{1-x} \, ,
\label{find1} \\
&&\mathcal{G}^{\hbox{\tiny LNC}}(x)=\frac{x}{8}\frac{1}{1-x} \, .
\label{find2}
\end{eqnarray}
They originate from the diagrams in the first and second raw of figure~\ref{fig:fig5} respectively.
The difference between the loop functions for the indirect case can be traced back to the different chiral projectors  sandwiching the intermediate composite neutrino propagator. Respectively a power of the intermediate, $M^*_J$, and incoming neutrino mass, $M_I^*,$ is selected in $\mathcal{G}^{\hbox{\tiny LNV}}(x)$ and $\mathcal{G}^{\hbox{\tiny LNC}}(x)$. The result in eqs.~(\ref{cp_direct}) and (\ref{cp_indirect}) have been obtained by summing the neutral and charged gauge bosons contributions to the widths that enter the CP asymmetry definition in (\ref{eq:adef}).

\subsection{Hierarchical and nearly degenerate limits}
The expressions for the flavoured direct and indirect CP asymmetry, given in (\ref{cp_direct}) and (\ref{cp_indirect}) respectively, stand for a quite general arrangement of the composite heavy-neutrino masses. It is useful to discuss two limiting cases often considered in leptogenesis: the hierarchical case, where $M^*_1 \equiv M^* \ll M^*_{i}$, with $i=2,3$, and the nearly degenerate case, where $M_2^* = M^* + \Delta $ with $0< \Delta  \ll M^*$. In order to keep the CP asymmetry expressions in a more compact form, we consider a sum over the final lepton flavour in the following. In the hierarchical case, starting from the CP asymmetries given in (\ref{cp_direct}) and (\ref{cp_indirect}), one can provide immediately both the direct and indirect contributions, whereas for the nearly degenerate case only the direct contribution can be derived straightforwardly. The indirect contribution will be obtained by taking into account a resummation in the heavy-neutrino propagators. 

Let us start with the hierarchical case. Here the lepton asymmetry is generated by the lightest composite neutrino because the heavier states decoupled from the dynamics much before.\footnote{This condition applies in standard leptogenesis with a seesaw type-I realization. It is often referred to as \textit{vanilla leptogenesis}~\cite{Blanchet:2012bk} where a strongly hierarchical spectrum and the unflavoured regime is considered.} It is then sufficient to study the evolution equations for the lightest neutrino number density and the corresponding lepton asymmetry induced in its LNV and CP violating decays. For a hierarchically ordered mass spectrum it is rather straightforward to obtain the expressions for the direct and indirect CP asymmetries. One has to perform an expansion in the small ratio $M^*/M^*_i$ of eqs.~(\ref{cp_direct}) and (\ref{cp_indirect}) and sum them up. The result reads, at leading order in $M^*/M^*_i$
\begin{equation}
\epsilon_{1} =  - \frac{5g^4+2g^2g'^2+g'^4}{3g^2+g'^2}  \left( \frac{M^*}{\Lambda}\right)^2  \sum_{i=2}^3  \frac{M^*}{M^*_i} \frac{ {\rm{Im}}\left[ (\tilde{f}_{1}^{*}\tilde{f}^{\phantom{*}}_{i})^2 \right] }{ 12 \pi|\tilde{f}_1|^2} + \dots \, ,
\label{limitCPhiera}
\end{equation}
where the dots stand for higher order terms in $(M^*/M^*_i)$. We note by passing that the term proportional to ${\rm{Im}}\left[ (\tilde{f}_{1}^{*}\tilde{f}^{\phantom{*}}_{i})(\tilde{f}^{\phantom{*}}_{\alpha 1}\tilde{f}^*_{\alpha i})  \right]$ in (\ref{cp_indirect}), that exactly vanishes in the unflavoured regime, is suppressed by an additional power of $M^*/M^*_i$. The direct and indirect contribution are of the same order of magnitude (we find $\epsilon_{1}^{\hbox{\tiny direct}}/\epsilon_{1}^{\hbox{\tiny indirect}} =-1/3 $, whereas in standard seesaw type I with right-handed neutrinos it holds $\epsilon_{1}^{\hbox{\tiny direct}}/\epsilon_{1}^{\hbox{\tiny indirect}}=1/2$~\cite{Covi:1996wh}).

On the other hand, the nearly degenerate case is more delicate and can lead to an interesting situation, i.e.~resonant leptogenesis~\cite{Flanz:1996fb,Pilaftsis:1997jf,Buchmuller:1997yu,Pilaftsis:2003gt,Garbrecht:2011aw,Garny:2011hg,Dev:2014laa}. In this case it is appropriate to distinguish between the direct and indirect contributions. We consider two heavy neutrino species in the following to present the results for the CP asymmetries.\footnote{On one hand the third heavy neutrino can be much heavier and decoupled already from the leptogenesis dynamics. On the other hand, we can consider an almost degenerate spectrum of the three composite neutrinos. We address both the cases in section~\ref{sec_out}.} We first provide the expression for the direct asymmetry. In this case the general result in eq.~(\ref{cp_direct}) has to be expanded for $\Delta  \ll M^*$. A finite splitting mass has to be kept, otherwise the asymmetry vanishes on general grounds: the CP phases can be rotated
away leading to purely real
effective couplings~\cite{Pilaftsis:1997jf}. The CP asymmetry for the composite neutrino of type 1 reads
\begin{equation}
\epsilon^{{\rm{direct}}}_{1}=   \frac{5g^4+2g^2g'^2+g'^4}{3g^2+g'^2} \left( \frac{M^*}{\Lambda}\right)^2 \frac{ {\rm{Im}}\left[ (\tilde{f}_{1}^{*}\tilde{f}^{\phantom{*}}_{2})^2  \right] }{64 \pi  |\tilde{f}_1|^2} \left[ 3-4 \ln 2 + \left( 11-16 \ln 2 \right)  \frac{\Delta}{M^*} \right] + \dots
\label{CP_dege_1}
\end{equation}
whereas for the neutrino of type 2 is
\begin{equation}
\epsilon^{{\rm{direct}}}_{2}= -  \frac{5g^4+2g^2g'^2+g'^4}{3g^2+g'^2} \left( \frac{M^*}{\Lambda}\right)^2 \frac{ {\rm{Im}}\left[ (\tilde{f}_{1}^{*}\tilde{f}^{\phantom{*}}_{2})^2 \right] }{64 \pi  |\tilde{f}_2|^2} \left[ 3-4 \ln 2 -\left( 5-8 \ln 2 \right)  \frac{\Delta}{M^*} \right] + \dots \, ,
\label{CP_dege_2}
\end{equation}
where the dots stand for higher order terms in the $\Delta/M^*$ expansion. The sum of eqs.~(\ref{CP_dege_1}) and (\ref{CP_dege_2}) does not vanishes in the limit  $\Delta \to 0$, however the sum of the width difference between decays into leptons and antileptons entering the numerator of (\ref{eq:adef}) does vanish. 

A rather different situation occurs for the indirect contribution. In this case the limit $\Delta \to 0$ does not provide a finite and meaningful result. This can be seen by eye looking at the structure of the functions $\mathcal{G}^{\hbox{\tiny LNV}}(x)$ and $\mathcal{G}^{\hbox{\tiny LNC}}(x)$ in eqs.~(\ref{find1}) and (\ref{find2}): they go to infinity when setting a vanishing mass splitting ($x \to 1$).  The problem is quite well known in the literature: the indirect contribution for nearly degenerate neutrinos can be understood as a mixing between the different neutrino species that makes the mass eigenstates different from the CP eigenstates~\cite{Flanz:1996fb}. Various approaches have been considered afterwards to address properly this situation~\cite{Pilaftsis:1997jf,Buchmuller:1997yu,Pilaftsis:2003gt,Garbrecht:2011aw,Garny:2011hg,Dev:2014laa}. The common idea is that the heavy neutrino may undergo many interactions
before decaying effectively into a lepton and a gauge boson pair (in the original formulation a lepton and Higgs boson pair). In doing so the intermediate neutrino acquires a finite width, and therefore its propagator is regulated for a vanishing mass splitting. One may also consider such resummation for the incoming neutrino (see e.g. \cite{Buchmuller:1997yu,Pilaftsis:2003gt}), however we introduce in the following the minimal condition to obtain a well-define result: only the width of the intermediate neutrino is accounted for. In doing so, we closely follow the original derivation given in Ref.~\cite{Pilaftsis:1997jf} for heavy Majorana neutrinos within seesaw type-I models and we obtain similar expressions for the CP asymmetries.

The indirect CP asymmetries in the degenerate case read
\begin{equation}
\epsilon^{{\rm{indirect}}}_{1} = -\frac{5g^4+2g^2g'^2+g'^4}{3g^2+g'^2} \left( \frac{M^*}{\Lambda}\right)^2 \frac{ {\rm{Im}}\left[ (\tilde{f}_{1}^{*}\tilde{f}^{\phantom{*}}_{2})^2)  \right] }{128 \pi |\tilde{f}_1|^2} \frac{M^* \Delta}{\Delta^2 + \Gamma_2^2/4} \, ,
\label{indiCPdege_1}
\end{equation}
and 
\begin{equation}
\epsilon^{{\rm{indirect}}}_{2} = -\frac{5g^4+2g^2g'^2+g'^4}{3g^2+g'^2} \left( \frac{M^*}{\Lambda}\right)^2 \frac{ {\rm{Im}}\left[ (\tilde{f}_{1}^{*}\tilde{f}^{\phantom{*}}_{2})^2  \right] }{128 \pi |\tilde{f}_2|^2} \frac{M^* \Delta}{\Delta^2 + \Gamma_1^2/4} \, ,
\label{indiCPdege_2}
\end{equation}
for the neutrino of type 1 and type 2 respectively, and the unflavoured widths $\Gamma_2$ and $\Gamma_1$ can be read off eq.~(\ref{gaugewidth_1}) when summing over $\alpha$. Now the vanishing mass splitting limit can be taken and one obtains a vanishing CP asymmetry as well~\cite{Pilaftsis:1997jf}. The expression of the CP asymmetries given in (\ref{indiCPdege_1}) and (\ref{indiCPdege_2}) provides an interesting speculation: requiring the condition $\Delta \sim \Gamma_1/2, \, \Gamma_2/2$ the CP asymmetry gets resonantly enhanced. The corresponding expression is
\begin{equation}
\epsilon_{1} = \epsilon_{2} \simeq -\frac{5g^4+2g^2g'^2+g'^4}{(3g^2+g'^2)^2} \frac{{\rm{Im}}\left[ (\tilde{f}_{1}^{*}\tilde{f}_{2}^{\phantom{*}})^2  \right] }{ 4 |\tilde{f}_1|^2 |\tilde{f}_2|^2} \, .
\label{CP_dege_res}
\end{equation} 
In the resonant case the CP asymmetry is not suppressed by the smallness of heavy-neutrino mass splitting, nor small ratios between
the heavy-neutrino masses nor the ratio $M^*/\Lambda$, the latter being typical of the model at hand.
\section{Out-of-equilibrium dynamics}
\label{sec_out}
We discuss here the third necessary condition for a successful leptogenesis, i.e. the out-of-equilibrium dynamics. For the sake of the present discussion we consider only heavy-neutrino decays and inverse decays~\cite{Buchmuller:2004nz,Buchmuller:2005eh}. The former process can induce a lepton asymmetry due to the LNV and CP violating decay process, whereas the latter can washout the asymmetry (a gauge boson and a lepton/antilepton combine to give a heavy neutrino). In order to observe a matter-antimatter asymmetry today, processes that washout the lepton asymmetry have to be inefficient at some epoch during leptogenesis. This usually happens when the temperature of the thermal medium drops below the mass of the heavy particle responsible for the generation of the matter-antimatter imbalance. 
In this regime inverse decays are Boltzmann suppressed ($\sim e^{-M^*/T}$) because SM gauge bosons and leptons/antileptons, which are kept in thermal equilibrium with the hot plasma, have typical energies much smaller than the heavy-composite neutrino masses. In other words they can hardly recombine into a massive state, with $M^* \gg T$, given that their typical energies is of order $T$. If it holds that heavy composite neutrinos are close to equilibrium till late times, namely $M^* \gg T$, then they can efficiently decay and generate a lepton asymmetry, whereas inverse decays are strongly suppressed.

\begin{figure}[t!]
\centering
\includegraphics[scale=0.54]{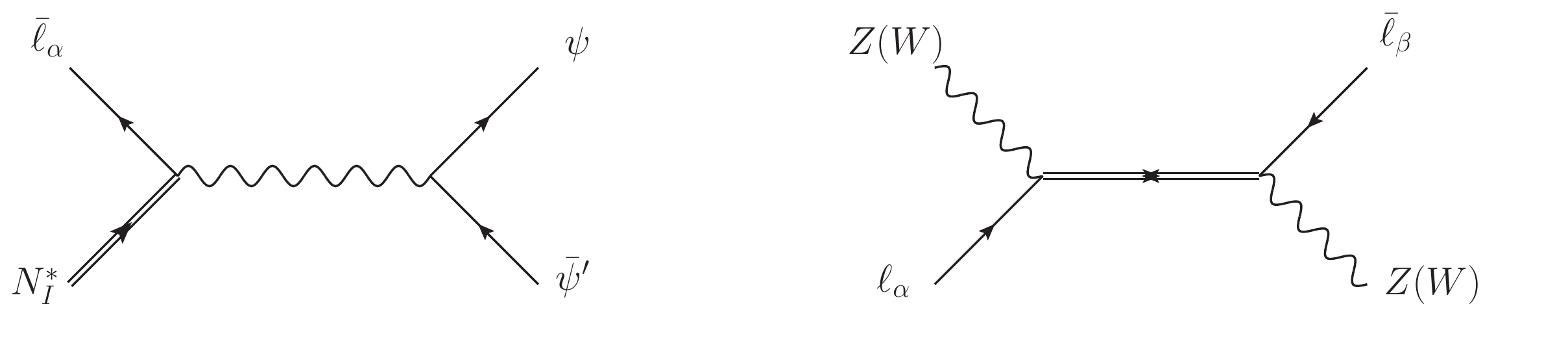}
\caption{\label{fig:fig6}Two examples for scattering diagrams involving a composite-heavy neutrino. They may be relevant for a full evaluation of the Boltzmann equations. Left: $\Delta L=1$ washout process that convert an antilepton-composite Majorana neutrino pair into a fermion-antifermion SM pair. Right: $\Delta L=2$ violating scattering mediated by a composite heavy neutrino.}
\end{figure}

There is a way to qualitatively study the out-of-equilibrium dynamics in leptogenesis in terms of the so-called  decay parameter~\cite{Buchmuller:2004nz,Davidson:2008bu}. This quantity is given by the ratio of the decay width of the heavy particle inducing a matter-antimatter asymmetry (here the composite neutrinos) and the Hubble rate. The former is evaluated at $T=0$ and we can then take the expressions in eqs.~(\ref{gaugewidth_1}) and (\ref{gaugecont_1}) for gauge and contact interactions respectively. The latter reads,
\begin{equation}
H=\sqrt{\frac{8 \pi^3 g_{\hbox{\tiny eff}}(T)}{90}} \frac{T^2}{M_{\hbox {\tiny Pl}}} \approx 1.66  \sqrt{g_{\hbox{\tiny eff}}} \frac{T^2}{M_{\hbox {\tiny Pl}}} \, ,
\label{hubble}
\end{equation}
and it is taken at temperatures of order of the heavy-particle mass, then $H(T=M^*)$ in our case. Then $g_{\hbox{\tiny eff}} \approx 100$ is the effective number of relativistic degrees of freedom at temperatures above the electroweak crossover (we take it as constant in the following estimations) and $M_{\hbox {\tiny Pl}}=1.2 \times 10^{19}$ GeV is the Planck mass. The decay parameter reads 
\begin{equation}
K_I=\frac{\Gamma_I}{H(T=M^*)}= 
\begin{cases}
\displaystyle \frac{|\tilde{f}_I|^2}{1.66 \sqrt{g_{\hbox{\tiny eff}}}} \frac{g'^2+3g^2}{32 \pi} \frac{M_{\hbox {\tiny Pl}}}{M^*_I} \left( \frac{M^*_I}{\Lambda} \right)^2 \, , 
\\ 
\displaystyle \frac{|\tilde{\eta}|^2}{1.66 \sqrt{g_{\hbox{\tiny eff}}}} \frac{g_*^4 \, 107}{1536 \pi^3} \frac{M_{\hbox {\tiny Pl}}}{M_I^*} \left( \frac{M^*_I}{\Lambda} \right)^4 \, .
\end{cases}
\label{decay_para}
\end{equation}
The out-of-equilibrium dynamics is normally established when the particle interaction rate, here measured by $\Gamma_I$, equals (and later on is smaller than) the universe expansion , i.~e.~interactions involving the composite neutrinos cannot catch up with the expansion of the universe. However a detailed analysis requires to study a set of rate equations, namely  Boltzmann equations with all the processes taken into account \cite{Davidson:2008bu,Kolb:1979qa,Nardi:2007jp}. The out-of-equilibrium dynamics is captured by the evolution of the neutrino and lepton-asymmetry number densities, which depend on input parameters like the heavy neutrino widths and the CP asymmetries, together with all the washout processes, i.e.~inverse decays and scatterings that work against the generation of a matter-antimatter asymmetry. It is beyond the scope of this work to solve the Boltzmann equations with all the relevant processes for the model at hand (see figure~\ref{fig:fig6} for an example of scattering processes). 

The decay parameter in (\ref{decay_para}) provides two limiting situations, called strong washout, $K_I \gg 1$  and weak washout $K_I \ll 1$, and they correspond to a close and far from equilibrium dynamics respectively. 
When the strong washout gets realized, the heavy states remain coupled with the hot plasma until late times, when the temperature drops below the mass scale $M^*_I$. Therefore the heavy composite neutrino enter a non-relativistic dynamics, being much heavier than the typical momentum scale of the heat bath, $T$, and inverse decays are Boltzmann suppressed. This ensures that the composite-neutrino decays effectively produce a different amount of leptons and antileptons. If one sticks to the choice made in the literature for the couplings $\eta$'s, $f$ and $f'$ in eqs.~(\ref{lag_mirr}) and (\ref{cont_2}) to be 1, and here it translates into $|\tilde{\eta}_I|^2 , |\tilde{f}_I|^2 \approx 1$, the condition $K_I \gg 1$ is realized. For example, for $M_I = 10$ TeV and $\Lambda=100$ TeV one obtains $K_I \sim \mathcal{O}(10^9)$  for both gauge and contact induced widths respectively. Hence there is room for a wide window of the effective couplings, namely $|\tilde{\eta}_I|^2, |\tilde{f}_I|^2 >10^{-9}$, that allows for late effective decays in a strong washout regime and hence ensuring the heavy neutrinos to be out of chemical equilibration. We notice that a similar result has been found in \cite{Zhuridov:2016xls} for the contact-interaction type Lagrangian.

\begin{figure}[t!]
\centering
\includegraphics[scale=0.82]{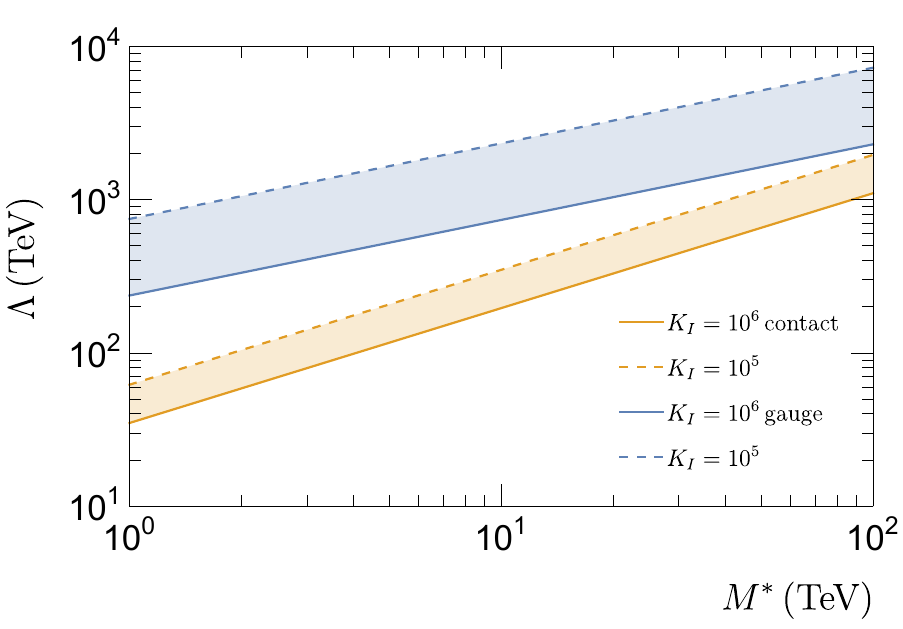}
\hspace{0.1 cm}
\includegraphics[scale=0.834]{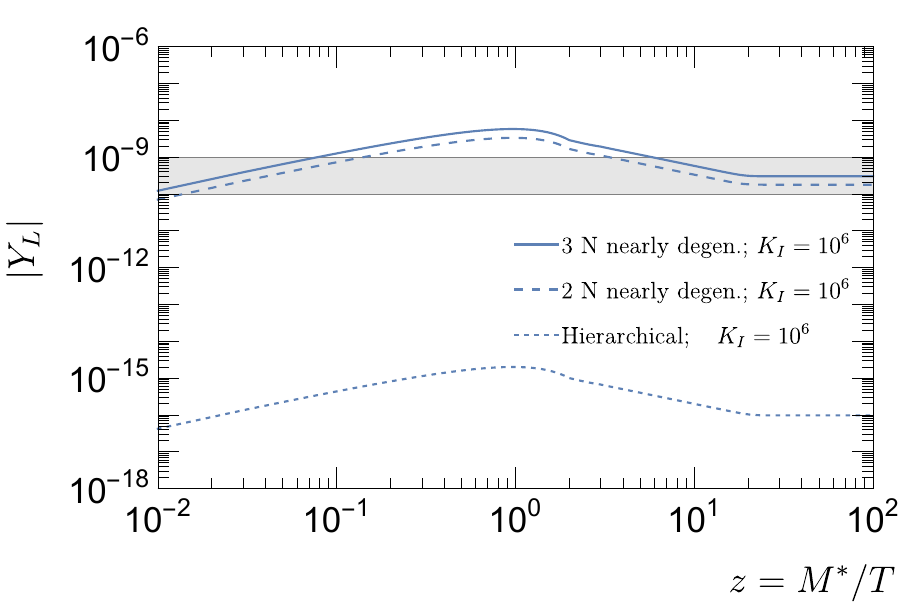}
\caption{\label{fig:fig7} Left: Contour level plots of the washout factors in the parameter space of the model $(\Lambda, M^*)$ for gauge and contact interactions, blue and orange band respectively. The effective couplings are fixed at $|\tilde{f}_I|^2= 10^{-2}=|\tilde{\eta}_I|^2= 10^{-2}$. Right: lepton asymmetry from Boltzmann equations. The decay parameter is fixed at $K=10^6$ and the effective couplings as for the left panel.  The solid and dashed blue lines correspond to two and three nearly degenerate neutrinos. The resonant conditions $\Delta =\Gamma_{I}/2$ and ${\rm{Im}}\left[ (\tilde{f}^*_I\tilde{f}^{\protect\phantom{*}}_J)^2  \right] /( |\tilde{f}_I|^2 |\tilde{f}_J|^2) \approx 1 $ imposed at the same time~\cite{Pilaftsis:1997jf}. The blue dotted-dashed line stands for the hierarchical case, with $M_1^*/M_i^{\protect\phantom{*}}=0.1$. The grey band originates from uncertainties in the value of the baryon asymmetry, efficiency factors converting the lepton asymmetry and neglected flavor effects.}
\end{figure}

As already mentioned, in order to be more quantitative one has to solve a network of Boltzmann equations. Few parameters have to be inserted into these equations, for example the decay parameter and the CP asymmetries provided within a given model~\cite{Kolb:1979qa,Buchmuller:2004nz}. Keeping the composite neutrino mass of order 1--10 TeV and taking the effective couplings, $|\tilde{f}_I|^2$ and $|\tilde{\eta}_I|^2$, not too small, the typical decay parameter for the model at hand is quite larger than the typical values considered in leptogenesis for the strong washout, namely $K \sim 10^0$--$10^3$~\cite{Buchmuller:2004nz,Buchmuller:2005eh}. This is mainly due to the appearance of the ratio $M^{\phantom{*}}_{{\rm{Pl}}}/M_{I}^*$, see (\ref{decay_para}). In figure \ref{fig:fig7} left, we show the contour levels for the washout factors from eq.~(\ref{decay_para}) for gauge and contact interactions. The blue and orange bands corresponds to the washout range $K_I \in \left[  10^5, 10^6\right] $, where the effective couplings are fixed such that $|\tilde{f}_I|^2 = |\tilde{\eta}_I|^2 = 10^{-2}$ (we do not consider smaller couplings in order not to spoil the descriptions in terms of effective operators whose coefficients are expected to be of $\mathcal{O}(f,f',\eta\text{'s}) \siml 1$). It is worth asking if we can still reproduce the order of magnitude of the baryon asymmetry for some choice of the parameters with very large washout factors. We perform an estimate as follows. We take $Y_B=(n_B-n_{\bar{b}})/s \approx 10^{-10}$~\cite{Larson:2010gs}, where $n_{B}(n_{\bar{B}})$ is the number density of baryons (antibaryons) and $s=h_{{\rm{eff}}}(2\pi^2/45)T^3$ is the entropy density of the universe.\footnote{Here $h_{{\rm{eff}}}$ are the relativistic degrees of freedom enetering the entropy density. This quantity is temperature dependent and differs (slightly) from $h_{{\rm{eff}}}$ in eq.~(\ref{hubble}).} Above the electroweak phase transition the sphaleron interactions convert approximately one third of the lepton asymmetry into a baryon one, therefore $Y_B \approx - Y_L/3$. We solve numerically the Boltzmann equations for $Y_L$ in a simplified scenario where only decays and inverse decays are considered~\cite{Pilaftsis:1997jf,Pilaftsis:1998pd,Buchmuller:2004nz,Buchmuller:2005eh}. For the network of Boltzmann equations we adopt the set up and formulation given in Ref.~\cite{Pilaftsis:1997jf,Pilaftsis:1998pd} for the rate equations. We assume equilibrium abundances for the composite neutrinos and a vanishing lepton asymmetry respectively at early times. 

As in standard thermal leptogenesis, we can classify different processes that contribute to the collision terms of the Boltzmann equations (see, e.g., \cite{Davidson:2008bu} for a detailed discussion). For the model at hand, we find three classes of processes that scale as $\tilde{f}^2 g^2$, $\tilde{f}^4 g^4$, $\tilde{f}^2 g^4$ (the same classification stands for $g \to \tilde{g}$). In this work we retain only processes of order $\tilde{f}^2 g^2$, for estimating the evolution of the lepton asymmetry.\footnote{Actually we have to work at the first non-trivial order $g^4\tilde{f}^4$ when calculating the numerator of the CP asymmetries as given in Eq.~(\ref{eq:adef}) since the corresponding width difference in the numerator vanishes at order $g^2\tilde{f}^2$.} The corresponding processes are decays, inverse decays and $s$-channel scatterings with on-shell contributions of intermediate heavy neutrinos, see diagram on the right of figure~\ref{fig:fig6}. When the on-shell region is met, the $2 \to 2$ process is of order $\tilde{f}^2 g^2$ instead of order $\tilde{f}^4 g^4$ as one would expect. This is well known in leptogenesis and it has been dubbed as Real Intermediate State (RIS) subtraction\cite{Kolb:1979qa}. Also for the model under study, the $s$-channel scattering process can be split into two terms: a first one that is of order $\tilde{f}^2 g^2$ (corresponding to the pole region) and a second one of order $\tilde{f}^4 g^4$ (away from the pole region). The former term, which we include in the numerical estimate, also ensures that the lepton asymmetry vanishes in equilibrium. 

The other scattering processes that contribute to washout terms of the Boltzmann equations are $t$- and $u$-channel $\Delta L=2$ scatterings mediated by heavy-composite neutrinos (and $\Delta L=2$ $s$-channel contribution after RIS subtraction), as well as $\Delta L=1$ scatterings involving a SM current, see figure~\ref{fig:fig6} left. They are of order $\tilde{f}^4 g^4$ and $\tilde{f}^2 g^4$ respectively. Since we take $\tilde{f} \approx 0.1$ and the SM gauge couplings are perturbative, i.e. smaller than one in the energies range of interest, we neglect these contributions in the following numerical estimate for the lepton asymmetry (the same approximation for the numerics has been adopted, e.g., in Ref.~\cite{Pilaftsis:1997jf,Pilaftsis:1998pd} and discussed in Ref.~\cite{Davidson:2008bu}).

For the gauge interactions, the result of the lepton asymmetry for both the hierarchical and nearly degenerate case is given in figure~\ref{fig:fig7} right.
In both cases we fix $|f_I|^2= 10^{-2}$, the gauge couplings are evolved at one-loop level in the SM, and the decay parameter is taken as $K_1 \equiv K =10^{6}$ (for the nearly degenerate case with two (three) neutrinos we take $K_{2(,3)} \equiv K$ as well). Moreover  we set $M_1^*/M^{\phantom{*}}_i=0.1$ with $i=2,3$ for the hierarchical case, whereas for nearly degenerate neutrino masses we impose the resonant condition $\Delta =\Gamma_{1,2}/2$, having then $\Delta/M^* \approx 10^{-8}$--$10^{-9}$ for $M_1^*=1$--$10$ TeV (similar splitting have been also obtained in \cite{Pilaftsis:1997jf,Pilaftsis:1998pd}). According to such a choice of the parameters the CP asymmetry is $\epsilon_{1} \approx 10^{-7}$ for the hierarchical case which, together with $K=10^{6}$, provides a final lepton asymmetry $|Y_L| \approx 10^{-16}$. This value is much smaller than the one observed, respectively, blue dotted-dashed line and grey band in figure~\ref{fig:fig7}. On the other hand, for the resonant case, we further require that ${\rm{Im}}\left[ (\tilde{f}_{1}^{*}\tilde{f}_{2}^{\phantom{*}})^2  \right] /( |\tilde{f}_1|^2 |\tilde{f}_2|^2) \approx 1$ to maximise the asymmetry in eq.~(\ref{CP_dege_res})~\cite{Pilaftsis:1997jf}. Therefore one obtains $\epsilon_{1} = \epsilon_{2} \simeq -(5g^4+2g^2g'^2+g'^4)/(4(3g^2+g'^2)^2) \approx 0.13$ for $M^* \approx 1$--$10$ TeV, the value of the heavy neutrino mass fixing those of the gauge couplings. In this case we find that there is room for a lepton asymmetry of the right order of magnitude, see solid blue line in figure~\ref{fig:fig7}, given that a fine tuning is implemented as described. We also consider the case of three nearly degenerate composite neutrinos similarly to standard leptogenesis~\cite{Casas:2001sr,Kageyama:2001zt,Pascoli:2003rq}. Indeed in the original formulation of the model at hand the composite neutrinos masses have been often taken to be the same. In order to keep the description simple we take $M_3 \equiv M^* +\Delta$, and then $M_2^*=M_3^*$ exactly. In so doing there is no contribution from the neutrino with mass $M_2$ to the CP violating decays of the neutrino with mass $M_3$ and viceversa, whereas both the heavier states contribute to the CP asymmetry of the lightest state (overall this results in the slight enhancement of the lepton asymmetry in figure~\ref{fig:fig7} right). The CP asymmetries for three nearly degenerate states are calculable from Ref.~\cite{Pilaftsis:2003gt}. 

\section{Conclusions}
\label{sec_conc}
In this paper we have studied the connection between composite models for fermions and leptogenesis. In particular we have considered the case of composite heavy Majorana neutrinos that are comprised in effective gauge and contact interaction Lagrangians. These heavy states are actively searched for at the LHC experiments. Majorana neutrinos are a key ingredient for the leptogenesis mechanism providing a source of lepton number violation. We have addressed the three Sakharov conditions for the model at hand, in order to check the minimal requirements for the generation of the matter-antimatter asymmetry. One important assumption has to be made: typical temperatures of the hot plasma have to be smaller than the compositeness scale so that composite neutrinos can contribute to the relevant degrees of freedom. Preons, or whatever the subconstituents may be called, are not manifest and leptogenesis is then driven by heavy Majorana neutrinos. Moreover one has to assume for the effective couplings, $\eta$'s, $f$ and $f'$ a complex nature, namely that they can develop non-vanishing complex phases. We have studied how composite models, addressing other fundamental questions, can work on the leptogenesis side with a focus on the CP asymmetry. We find that leptogenesis can be implemented in composite neutrino models described by the effective Lagrangians (\ref{eff_lag_gau}) and (\ref{eff_lag_cont}).

In the case of gauge interactions, we have calculated explicitly the CP violating parameters, defined in (\ref{eq:adef}), induced by heavy composite neutrino decays. This is our main original result on the computational side. In a mass arrangement that reads $M^*_1 < M^*_2 < M^*_3$, we derived the direct and indirect contribution to the CP asymmetry. The corresponding results read off eq.~(\ref{cp_direct}) and eq.~(\ref{cp_indirect}) respectively. The one-loop diagram inducing the direct contribution is more involved than the one in standard leptogenesis due to a more complicated Dirac algebra and additional powers of the loop momentum, given in turn by the magnetic-type interaction (see eq.(\ref{eff_lag_gau})). The indirect contribution instead resembles more closely the standard result in the literature. The suppression ratio $M^*_I/ \Lambda$ appears in the expressions of the CP asymmetries as inherited from the widths. Then we have provided the CP asymmetries in the strongly hierarchical and nearly degenerate limits, including also the case of resonant leptogenesis. The phenomenological impact of a resonantly enhancement is really important for the standard seesaw type I leptogenesis. Indeed it allows to low  the heavy neutrino mass scale down to the TeV scale. Here we do not gain much in this respect, as the mass of the heavy composite neutrinos is expected to be of the order of the TeV scale from the original set up of the model and LHC direct searches~\cite{CMSexp,Sirunyan:2017xnz}. However, in the resonant case, the absence of the suppression factor, $(M^*/\Lambda)^2$,    has a numerically important impact on the lepton asymmetry from composite neutrino decays.
Indeed we solved the Boltzmann equations in the simplified scenario where only decays and inverse decays are considered. When keeping the composite neutrino mass of order $M^*\sim1-10$ TeV  and couplings $f_I \sim 0.1$, only the resonantly enhanced CP asymmetry is enough to reproduce the correct order of magnitude for the lepton asymmetry. A very large washout and small CP asymmetries prevents the same to occur in the hierarchical case. We stress that we have not taken into account a full and systematic treatment in terms of Boltzmann equations: only the leading order processes, namely those at order $\tilde{f}^2 g^2$, have been considered. These are decays, invese decays and $s$-channel scatterings with on-shell intermediate composite neutrinos.  Flavour effects were not considered as well. Moreover, in order to properly handle  the saturation of the resonant enhancement for $\Delta \siml \Gamma_I/2$, it would be necessary to include also coherent transition between the Majorana neutrino states~\cite{Garny:2011hg,Garbrecht:2011aw,Dev:2014laa,Dev:2014wsa}. However this is beyond the accuracy of our work.

As far as leptogenesis induced by gauge interactions is concerned, the phase space parameter compatible with a successful generation of the matter-antimatter asymmetry is likely out of reach of present-day colliders. This is due to the large values of the compositness scale, $\Lambda \approx 10^2$ TeV for $M^* \approx 1$ TeV, together with smaller effective couplings as usually taken in the experimental analysis. Correspondingly production cross sections would be much suppressed and decays into SM particles would as well.    

We foresee at least two future directions for further developments of the topics here discussed. First, the lepton asymmetry induced by contact interactions should be considered and included in the leptogenesis dynamics. Contact interactions are indeed an important ingredient in experimental analyses and corresponding searches of composite neutrino states. The CP asymmetries are not known for the model studied in this paper, though partial results are available in the literature that refer to models with similar effective operators as in eq.~(\ref{cont_2})~\cite{Zhuridov:2016xls}. In the model considered in this work, contact interactions provide much smaller values for $\Lambda$, when $M^*$ is taken in the TeV range, compared to what happens with  gauge interactions (see Fig.~\ref{fig:fig7}, left panel, lower-orange band). Then it is worth addressing the CP asymmetries induced by contact interactions and the corresponding lepton asymmetries in the near future~\cite{BP2}. 
Second, a complete study of the Boltzmann equations to asses quantitatively the leptogenesis mechanism within composite neutrino models is on order. The full set of processes entering the collision terms of the Boltzmann equations for both gauge and contact interactions has to be considered for a complete and conclusive analysis. In doing so one may give more precise benchmarks on the mass scale, $M^*$, the compositeness scale, $\Lambda$, and effective couplings, $f'$s and $\eta$'s, that reproduce the correct matter-antimatter asymmetry. One could then relate the parameter space compatible with a successful leptogenesis to the present and predicted exclusion bounds at the LHC (and possibly future colliders) for composite neutrino models in terms of the very same parameters. 
\subsection*{Acknowledgements}
The work of S.\ B. was partly supported by the Swiss National Science Foundation (SNF) under grant 200020-168988.
\appendix \markboth{Appendix}{Appendix}
\renewcommand{\thesection}{\Alph{section}}
\numberwithin{equation}{section}
\section{Loop integrals}
\label{app1}
In the language adopted in the paper, the width can be traced back to self-energy diagrams for the heavy composite neutrinos. In the computation of the CP asymmetry the decay processes into lepton and antileptons have to be disentangled. This amounts at properly cutting the two-loop self-energy diagrams through a lepton or an antilepton line.  The remaining one-loop amplitude has to be evaluated and only its imaginary part contribute to the asymmetry. Indeed the real parts from the lepton and antilepton contribution cancel in the numerator of the CP asymmetry (\ref{eq:adef}). In the following we show explicitly the  vertex and wave-function one-loop integrals that one has to evaluate to obtain the results in eqs.~(\ref{direct_loop}), (\ref{find1}) and (\ref{find2}). We denote with the generation index $I$ the incoming external neutrino, whereas we label the internal one with $J$.
\subsection{Vertex diagram}
We start with the vertex diagram, the loop integral reads
\begin{equation}
\mathcal{I}_{{\rm{direct}}}=\mathcal{T}^{\lambda \tau \rho} F_{\lambda \tau \rho} \, ,
\end{equation}
where the Dirac structure and one-loop momentum integral are respectively
\begin{eqnarray}
\mathcal{T}^{\lambda \tau \rho} &=& M^*_J (p-q)_\mu(p-q)_{\omega} \sigma^{\mu \nu}  \slashed{q} \sigma^{\lambda \eta} \sigma^{\omega}_{\: \nu} \gamma^\tau \sigma^{\rho}_{\: \eta} \, ,
\label{Dir_vert}
\end{eqnarray}  
\begin{equation}
F_{\lambda \tau \rho} =\int \frac{d^4 \ell}{(2 \pi)^4} \frac{i^3 \ell_{\lambda} (p-\ell)_{\tau} \ell_{\rho} }{\left[ \ell^2 +i\epsilon\right] \left[ (p-\ell)^2  +i\epsilon \right]  [\left( \ell -q \right)^2 -M^{*2}_J +i\epsilon ]  } \, .
\label{loop_vert}
\end{equation} 
Here $q^{\mu}$ and $(p-q)^\mu$ are the lepton/antilepton and gauge boson momenta respectively and we cut the corresponding propagators in the two-loop diagrams in figure~\ref{fig:fig4}, whereas $\ell^{\mu}$ denote the remaining one-loop momenta. 
The heavy composite neutrino masses, $M_I^*$ and $M_J^*$, are the only mass scales appearing in the loop since the SM particles are taken as massless. The incoming heavy composite neutrino momentum is $p^{\mu}$. We perform the calculation by imposing $p^{\mu}=M_I^*v^\mu$ with $v^{\mu}=(1,\bm{0})$, and therefore $q^{\mu}=M_I^*/2u^{\mu}$ with $u^2=0$ after the cut. We remark that there are two additional powers of the loop momentum with respect to the vertex integral in standard leptogenesis with right-handed neutrinos~\cite{Covi:1996wh}. The origin is the derivative-type coupling in the Lagrangian~(\ref{eff_lag_gau}).  The integral can be carried out with standard one-loop techniques and we need only the imaginary part of the integral (\ref{loop_vert}) to obtain the CP asymmetry.
\subsection{Wave-function diagram}  
Now we discuss the direct contribution to the CP asymmetry. Here we have to split the discussion according to the two sets of two-loop diagrams in the first and second raw of figure~\ref{fig:fig5}. Starting with the diagrams in the first raw, the remaining one-loop integral after the cut reads
\begin{equation}
\mathcal{I}_{{\rm{indirect}}}=\mathcal{U}^{\lambda \tau \rho} G_{\lambda \tau \rho} \, ,
\end{equation}
where 
\begin{equation}
\mathcal{U}^{\lambda \tau \rho}= \frac{i M_{J}}{p^2-M_J^2} \, \sigma^{\lambda \omega} \gamma^{\tau} \sigma^{\rho}_{\:\: \omega} \, ,
\label{Dir_wave}
\end{equation}
\begin{equation}
G_{\lambda \tau \rho} =\int \frac{d^4 \ell}{(2 \pi)^4} \frac{i^2 \ell_{\lambda} (\ell-p)_{\tau} \ell_{\rho} }{\left[ \ell^2 +i\epsilon\right] \left[ (p-\ell)^2  +i\epsilon \right]    } \, .
\label{loop_wave}
\end{equation} 
Here we are interested in the real part of the loop integral because, even though the intermediate neutrino propagator contributes to the loop amplitude, it does not appear in the loop integral (see (\ref{Dir_wave})). Therefore the only mass scale that appears in (\ref{loop_wave}) is the incoming neutrino mass at variance with the vertex integral where both the incoming and internal heavy-neutrino mass play a role. Finally there is a third quantity to consider, namely the one-loop amplitude when the sum over the final lepton/antilepton flavor is not performed. In this case the diagram $d$ in figure~\ref{fig:fig3} also contributes, that provides the two-loop amplitude in the second raw of figure~\ref{fig:fig5}. The corresponding one-loop integral reads
\begin{equation}
\mathcal{I}_{{\rm{indirect}}}^{{\rm{flavor}}}=\mathcal{V}^{\lambda \tau \rho} G_{\lambda \tau \rho} \, ,
\label{int_fla}
\end{equation}
where 
\begin{equation}
\mathcal{V}^{\lambda \tau \rho}= \frac{i \slashed{p}}{p^2-M_J^2} \, \sigma^{\lambda \omega} \gamma^{\tau} \sigma^{\rho}_{\:\: \omega} \, ,
\label{Dir_wave_2}
\end{equation}
and the one-loop integral in (\ref{int_fla}) is again that given in (\ref{loop_wave}).

\end{document}